\documentclass[12pt]{iopart}

\usepackage[dvipdfmx]{graphicx}

\usepackage{cases}
\usepackage{bm}
\usepackage{amssymb}
\usepackage{amsthm}
\usepackage{braket}

\usepackage{cite}

\begin{document}

\title[Dissipation Effect on Optical Force and Torque near Interfaces]{Dissipation Effect on Optical Force and Torque near Interfaces}

\author{Daigo Oue \footnote{Present address: Department of Physics, Imperial College London
Prince Consort Road, Kensington, London SW7 2AZ, UK.} }

\address{Division of Frontier Materials Science, Osaka University, 1-3 Machikaneyama, Toyonaka, Osaka, Japan 560-8531}
\ead{daigo.oue@gmail.com}
\vspace{10pt}
\begin{indented}
\item[]March 2019
\end{indented}

\begin{abstract}
The Fresnel-Snell law, which is one of the fundamental laws in optics and gives insights on the behaviour of light at interfaces, is violated if there exists dissipation in the transmitting media. 
In order to overcome this problem, we extend the angle of refraction from a real number to a complex number. 
We use this complex-angle approach to analyse the behaviour of light at interfaces between lossy media and lossless media. 
We reveal that dissipation makes the wavenumber of the light exceed the maximum allowed at lossless interfaces. 
This is surprising because, in general, dielectric loss only change the intensity profiles of the light, so this excess wavenumber cannot be produced in the bulk even if there exists dielectric loss. 
Additionally, anomalous circular polarisation emerges with dissipation. 
The direction of the anomalous circular polarisation is transverse, whereas without dissipation the direction of circular polarisation has to be longitudinal. 
We also discuss how the excess wavenumber can increase optical force and how the anomalous circular polarisation can generate optical transverse torque.
This novel state of light produced by dissipation will pave the way for a new generation of optical trapping and manipulation.
\end{abstract}

%
%
%
%
%

\section{Introduction}
Optical force and optical linear momentum were theoretically predicted by Maxwell \cite{maxwell1865viii} and observed by Nichols and Hull \cite{nichols1903pressure, nichols1903pressure2nd}. 
The existence of optical torque and optical angular momentum was theoretically proposed by Poynting \cite{poynting1909wave}, and experimentally confirmed by Beth \cite{beth1936mechanical}.
Although the interesting fact that light can push and rotate objects attracted many scientists, not much progress was made in the study of optical force and torque until the availability of a high-intensity light source for generating large forces or torques. 
Around 1950, the laser was invented in Bell laboratories. 
Ashkin, who also belonged to Bell laboratories at that time, demonstrated manipulation of microparticles with laser beam \cite{ashkin1970acceleration, ashkin1986observation}.
After Ashkin's experiments, many interesting setups for optical manipulation have been proposed.
Some of them function in the optical near field using evanescent waves and surface plasmon polaritons \cite{kawata1992movement, kawata1996optically, okamoto1999radiation, volpe2006surface, Hassanzadeh2014, Hassanzadeh2015, Hassanzadeh2016, Mohammadnezhad2017}.
The optical near field has peculiar properties such as steep intensity gradients, large wavevectors compared to propagating field, and transverse circular polarisation.
There have been many studies of optical near field forces and torques \cite{chaumet2000coupled, canaguier2013force, canaguier2014transverse, bliokh2014extraordinary, antognozzi2016direct}, but the effect of dissipation, i.e. dielectric or magnetic loss, especially near interfaces, has never been discussed before.
Electromagnetic fields also decay due to dissipation (without any total internal reflection), and in dissipative media they can have near-field-like characteristics such as intensity gradients.
In this paper, we use a complex-angle approach to analyse the behaviour of optical field near interfaces with dissipation, and analyse how the dissipation contributes to optical force and torque.
\section{Electromagnetic fields near interfaces with dissipation}
\begin{figure}[htbp]
\begin{center}
  \includegraphics[width=8cm]{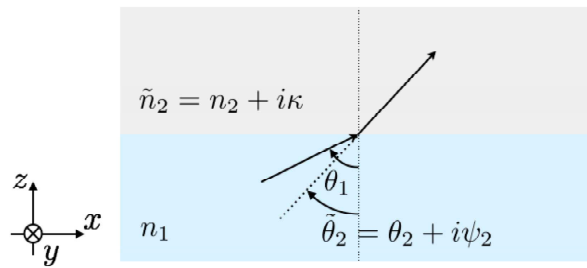}
  \caption{
    Refraction of light at interface.
    We consider light incidence on interface between non-dissipative media ($n_1$) and dissipative media ($\tilde{n}_2 = n_2 + i\kappa_2$). 
    Let the angle of incidence $\theta_1$ and the angle of refraction $\tilde{\theta}_2$.
  }
  \label{fig:configuration}
\end{center}
\end{figure}

In order to calculate the electromagnetic field near dissipative interfaces, we further develop a complex-angle approach, which is originally proposed by Bekshaev \textit{et. al.} to deal with total internal reflection \cite{bekshaev2013mie}.
FIG. \ref{fig:configuration} shows the situation we consider in this paper.
We consider interfaces between dissipative media and non-dissipative media.
The field is incident on the interface from the lower side, and we take $\theta_1$ for the angle of incidence.
In dissipative media, the refractive index must be a complex value with non-zero imaginary part $\left(\tilde{n}_2 \in \mathbb{C}\right)$.
\begin{equation}
  \tilde{n}_2 = n_2 + i\kappa_2.
\end{equation}
Considering Snell's law,
\begin{equation}
  n_1 \sin \theta_1 = \tilde{n}_2 \sin \tilde{\theta}_2. \label{eq:snell_law_dissipative}
\end{equation}
In this equation, the left hand side is real ($n_1,\sin \theta_1 \in \mathbb{R}$), while $\tilde{n}_2$ has imaginary part.
Thus, $\sin \tilde{\theta}_2$ is not a real number any longer, but it is a complex number with non-zero imaginary part $\left(\sin \tilde{\theta}_2 \in \mathbb{C} \right)$, and the angle of refraction must be a complex number with non-zero imaginary part,
\begin{equation}
  \tilde{\theta}_2 = \theta_2 + i\psi_2 (\psi < 0),
\end{equation}
which satisfies the following simultaneous equations (modified Snell's law):
\begin{subnumcases}
  {}\mathrm{Re}(\tilde{n}_2 \sin \tilde{\theta}_2) = n_1 \sin \theta_1, \label{eq:theta_diss_cond1}\\
  \mathrm{Im}(\tilde{n}_2 \sin \tilde{\theta}_2) = 0. \label{eq:theta_diss_cond2}
\end{subnumcases}

\par Once we obtain the angle of refraction, we can calculate the explicit expressions of the transmitted field by rotating $+z$-propagating plane wave $\bm{E} = \bm{E}_0 \exp (i\bm{k} \cdot \bm{r})$ towards the directions of refraction. 
Here,
\begin{eqnarray}
  \bm{E}_0 = \left(\begin{array}{c}E_p\\ E_s\\ 0 \end{array}\right), 
  \bm{k} = \left(\begin{array}{c}0\\ 0\\ \tilde{n}_2 k_0 \end{array}\right).
\end{eqnarray}
$k_0$ is the wavenumber in vacuum.
For the rotation, we use
\begin{equation}
  R (\theta) = \left(\begin{array}{ccc}
    \cos \theta & 0 & \sin \theta \\
    0 & 1 & 0 \\
    -\sin \theta & 0 & \cos \theta
  \end{array}\right).
\end{equation}

\par By substituting the complex-angle of refraction $\tilde{\theta}_2$ into the rotation matrix, we get a complex-angle rotation matrix which gives the wavevector of electromagnetic field in dissipative media:
\begin{eqnarray}
  \bm{k} &\rightarrow \tilde{\bm{k}} = R(\tilde{\theta}_2)\bm{k} = \bm{k} + i\bm{\eta}, \label{eq:wavevector_diss_1}
\end{eqnarray}
where
\begin{eqnarray}
  \bm{k} &= k_0\left(\begin{array}{c}
    n_2 \sin \theta_2 \cosh \psi_2 - \kappa_2 \cos \theta_2 \sinh \psi_2 \\
    0\\
    n_2 \cos \theta_2 \cosh \psi_2 + \kappa_2 \sin \theta_2 \sinh \psi_2
  \end{array}\right), \label{eq:wavevector_diss_2}\\
  \bm{\eta} &= k_0\left(\begin{array}{c}
    0\\
    0\\
    -n_2 \sin \theta_2 \sinh \psi_2 + \kappa_2 \cos \theta_2 \cosh \psi_2
  \end{array}\right). \label{eq:wavevector_diss_3}
\end{eqnarray}
The imaginary part $\bm{\eta}$ represents the decay of the field, and the real part $\bm{k}$ represents the propagation of the field.
Note that, from the second equation of the modified Snell's law,
\begin{eqnarray}
 \eta_x \propto n_2 \cos \theta_2 \sinh \psi_2 + \kappa_2 \sin \theta_2 \cosh \psi_2 = 0.
\end{eqnarray}

\par
\begin{figure}[tbp]
  \centering
  \includegraphics[width=8cm]{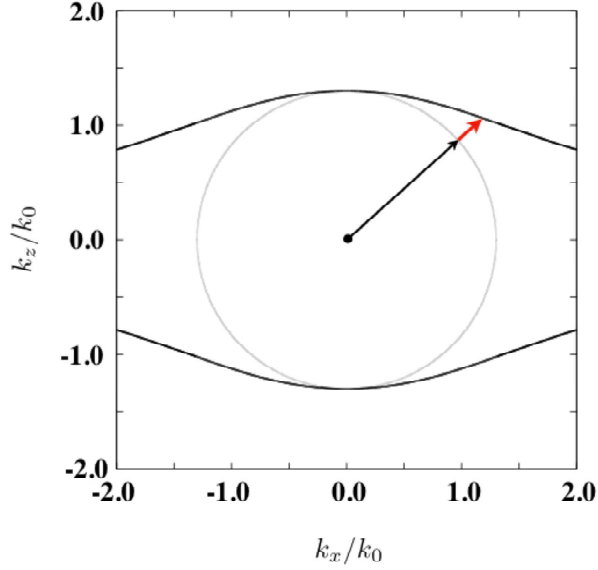}
  \caption{
    Excess wavevector is produced by dissipation.
    The grey curve is the iso-frequency curve without the presence of dissipation ($\tilde{n}_2=1.3+0i$), while the black solid curve is the iso-frequency with the presence of dissipation ($\tilde{n}_2=1.3+1.3i$).
  }
  \label{fig:excess}
\end{figure}

 FIG. \ref{fig:excess} shows the iso-frequency curves followed by the real part of the wavevector for fixed $k_0$ near the interface between a dissipative medium and various lossless media.
The solid black curve is the curve with the presence of dissipation ($\tilde{n}_2 = 1.3 + 1.3i$), and the grey curve is without dissipation ($\tilde{n}_2 = 1.3 + 0i$).
It can be said that the curve with dissipation is greater than the non-dissipative curve, which means that higher (spatial) frequency field is produced by dissipation and that this could enlarge the optical force on particle in the field.
More excess wavenumber is produced for larger angle of incidence.
Physically, this excess wavenumber in the $x$-direction is produced at the expense of the wavenumber in the $z$-direction.

\par By the complex-angle rotation, we can also calculate the electric field vector,
\begin{eqnarray}
  \bm{E}_0 \rightarrow \tilde{\bm{E}}_0 = R(\tilde{\theta}_2) \left(\begin{array}{c} \mathcal{T}_p E_p \\ \mathcal{T}_s E_s \\ 0 \end{array} \right) \\
           = \left(\begin{array}{c} \mathcal{T}_p E_p (\cos \theta_2 \cosh \psi_2 - i\sin \theta_2 \sinh \psi_2 ) \\ \mathcal{T}_s E_s\\ -\mathcal{T}_p E_p (\sin \theta_2 \cosh \psi_2 + i\cos \theta_2 \sinh \psi_2) \end{array}\right).
\end{eqnarray}
Here, $\mathcal{T}_{p}$ and $\mathcal{T}_{s}$ are Fresnel coefficients.
Finally, we get the explicit representation of the refracted field near an interface with dissipation:
\begin{eqnarray}
  \tilde{\bm{E}} &= \tilde{\bm{E}}_0 \exp \left(i\tilde{\bm{k}}\cdot \bm{r}\right) \\
                 &= \tilde{\bm{E}}_0 \exp \left(i \bm{k}\cdot \bm{r}\right) \cdot \exp \left(- \bm{\eta} \cdot \bm{r}\right). \label{eq:E_wave_dissipation}
\end{eqnarray}
In this paper, we are interested in p-polarisation since it has non-trivial longitudinal field vector, so from here we set $E_s = 0$ and for simplicity set $E_p = 1$ and $n_1 = 1$, but it is straightforward to discuss the general case ($E_s, E_p \neq 0$, and $n_1 \geq 1$) using our approach. Note that (\ref{eq:wavevector_diss_2}) to (\ref{eq:E_wave_dissipation}) can also be derived directly from Maxwell's equations.

\par
\begin{figure}[tbp]
  \centering
  \includegraphics[width=10cm]{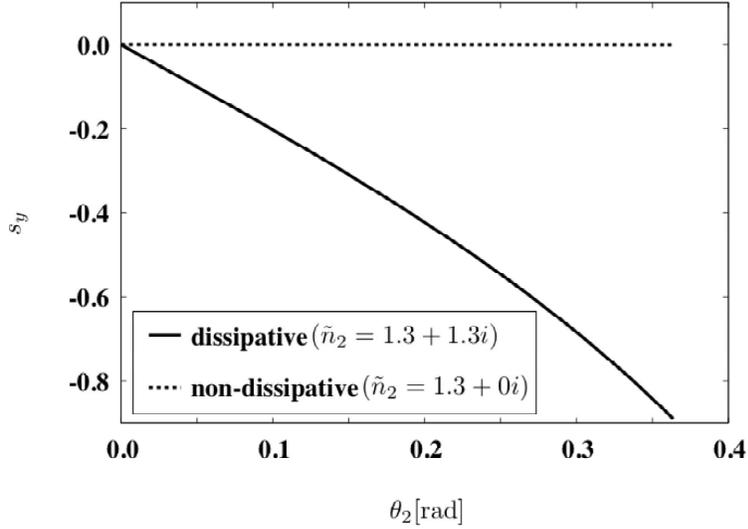}
  \caption{
    Anomalous circular polarisation (transverse $y$ component of $\bm{s}$) emerges with dissipation ($\tilde{n_2}=1.3+1.3i$).
  }
  \label{fig:anomalous}
\end{figure}

 We can see that there is phase difference between the transverse $z$ component of the field vector and the longitudinal $x$ component.
That is because the continuity conditions for the tangential component of field and for the normal component are different, and this induces phase difference between the two components, and causes rotation of the field.
To describe the degree of circular polarisation and direction of the field rotation, we can use a psuedovector
\begin{eqnarray}
  \bm{s} = \frac{g}{4\omega} \mathrm{Im} \left( \bm{E}^* \times \bm{E} \right),
\end{eqnarray}
where $g = 1/4\pi$ is a Gaussian-unit factor.
For our field, we have
\begin{eqnarray}
  \bm{s} = \frac{g}{2\omega}\sinh \psi_2 \cosh \psi_2 \exp \left(-2\bm{\eta} \cdot \bm{r}\right) \bm{e}_y.
\end{eqnarray}
Here, $\bm{e}_y$ is the unit vector in the direction of $y$.
We can observe if we flip the sign of $\theta_2$, then, from (\ref{eq:theta_diss_cond2}), the sign of $\psi_2$ is also flipped, and thus the direction of the psuedovector $\bm{s}$ is flipped.
This is one kind of spin-momentum locking \cite{van2016universal}, in which the direction of circular polarisation and that of propagation are tied to each other. 
In FIG. \ref{fig:anomalous}, we plot the transverse $y$ component of the psuedovector.
This implies that the field is rotating in the transverse direction and that we can exert transverse optical torque on a particle in the field.

\section{Optical force and torque near dissipative interfaces}
\begin{figure}[htbp]
  \centering
  \includegraphics[width=8cm]{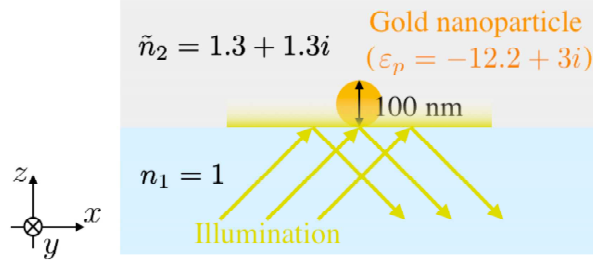}
  \caption{
    Schematic image of the setup for calculating optical force and torque. 
    We have a small probe particle on the interface between dissipative media and transparent media. 
    The diameter of the particle is 100 nm.
  }
  \label{fig:exp_setup}
\end{figure}

In this section, we discuss the effect of dissipation on optical forces and torques.
We consider the situation shown in FIG. \ref{fig:exp_setup}.
There is a small probe particle on the interface between dissipative media and lossless transparent media.
For the calculation of the optical force and torque, we assume that the particle is small enough compared to the wavelength of the field so that it can be regarded as a point dipole.
The time averaged optical force and torque exerted on a point dipole is given by\cite{nieto2010optical, bekshaev2013subwavelength}
\begin{eqnarray}
  \bm{F} &= \Braket{(\mathfrak{d}_e \cdot \nabla)\mathcal{E} + (\mathfrak{d}_m \cdot \nabla)\mathcal{H} + \frac{1}{c} \dot{\mathfrak{d}_e} \times \mathcal{B} - \frac{1}{c} \dot{\mathfrak{d}_m} \times \mathcal{D}} 
  \propto \mathrm{Im}(\alpha_e) \bm{k},\label{eq:opt_force}\\
  \bm{T} &= \Braket{\mathfrak{d}_e \times \mathcal{E} + \mathfrak{d}_m \times \mathcal{H}} 
  \propto \mathrm{Im}(\alpha_e) \bm{s}.\label{eq:opt_torque}
\end{eqnarray}
Here, $\mathcal{E}$ and $\mathcal{H}$ are real fields associated with complex fields, $\mathcal{E} = \mathrm{Re}\left(\bm{E}e^{-i\omega t}\right)$ and $\mathcal{H} = \mathrm{Re} \left(\bm{H}e^{-i\omega t}\right)$. We do not consider higher order contributions, which could cause electric-magnetic dipolar interaction force \cite{nieto2010optical} and torque \cite{nieto2015optlett, nieto2015pra}, but focus on the fundamental order by the dipole approximation.
These complex fields satisfy the monochromatic Maxwell's equations:
\begin{eqnarray}
  \nabla \cdot \bm{E} = \nabla \cdot \bm{H} = 0,\\
  \nabla \times \bm{E} = i\frac{\omega}{c}\mu \bm{H},\\
  \nabla \times \bm{H} = -i\frac{\omega}{c}\varepsilon \bm{E}.
\end{eqnarray}
Complex flux densities, $\bm{D}$ and $\bm{B}$, are characterized by permittivity $\varepsilon$ and permeability $\mu$: $\bm{D}= \varepsilon \bm{E}$, $\bm{B} = \mu \bm{H}$, and these give the real fields: $\mathcal{D} = \mathrm{Re}\left(\bm{D}e^{-i\omega t}\right)$, $\mathcal{B} = \mathrm{Re}\left(\bm{B}e^{-i\omega t}\right)$.
We also use complex dipole moments $\bm{d}_e$, $\bm{d}_m$ to give real dipole moments: $\mathfrak{\bm{d}}_{e}(\bm{r},t) = \mathrm{Re} \left(\bm{d}_{e}(\bm{r})e^{-i\omega t}\right)$, $\mathfrak{\bm{d}}_{m}(\bm{r},t) = \mathrm{Re}\left(\bm{d}_{m}(\bm{r})e^{-i\omega t}\right)$.
Electric and magnetic dipoles are characterized by electric polarisability $\alpha_e$ and by magnetic polarisability $\alpha_m$: $\bm{d}_e = \alpha_e \bm{E}$, $\bm{d}_m = \alpha_m \bm{H}$.

\begin{figure}[tbp]
  \begin{minipage}[t]{0.5\hsize}
    \centering
    \includegraphics[width=8cm]{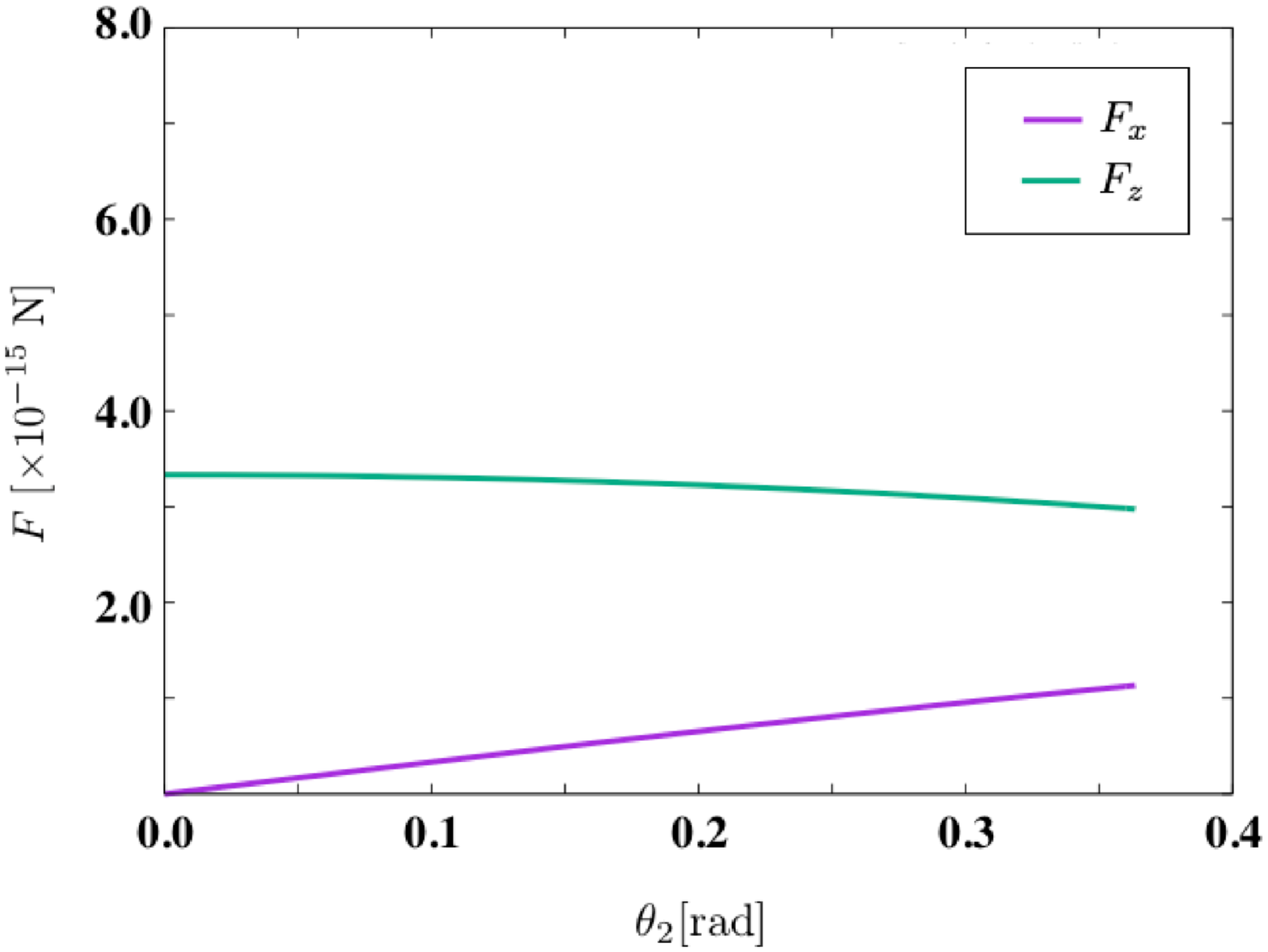}
  \end{minipage}
  \begin{minipage}[t]{0.5\hsize}
    \centering
    \includegraphics[width=8cm]{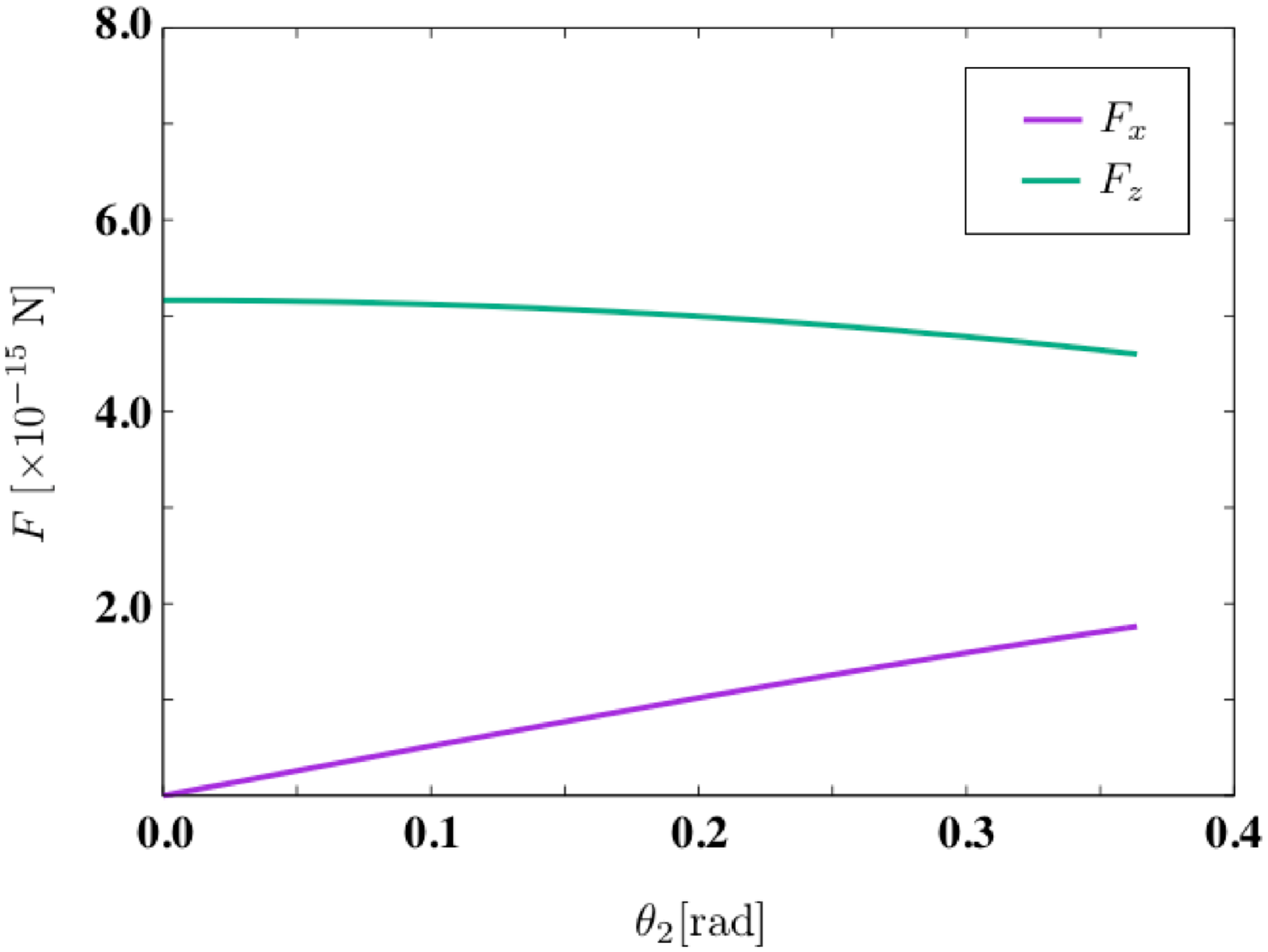}
    \includegraphics[width=8cm]{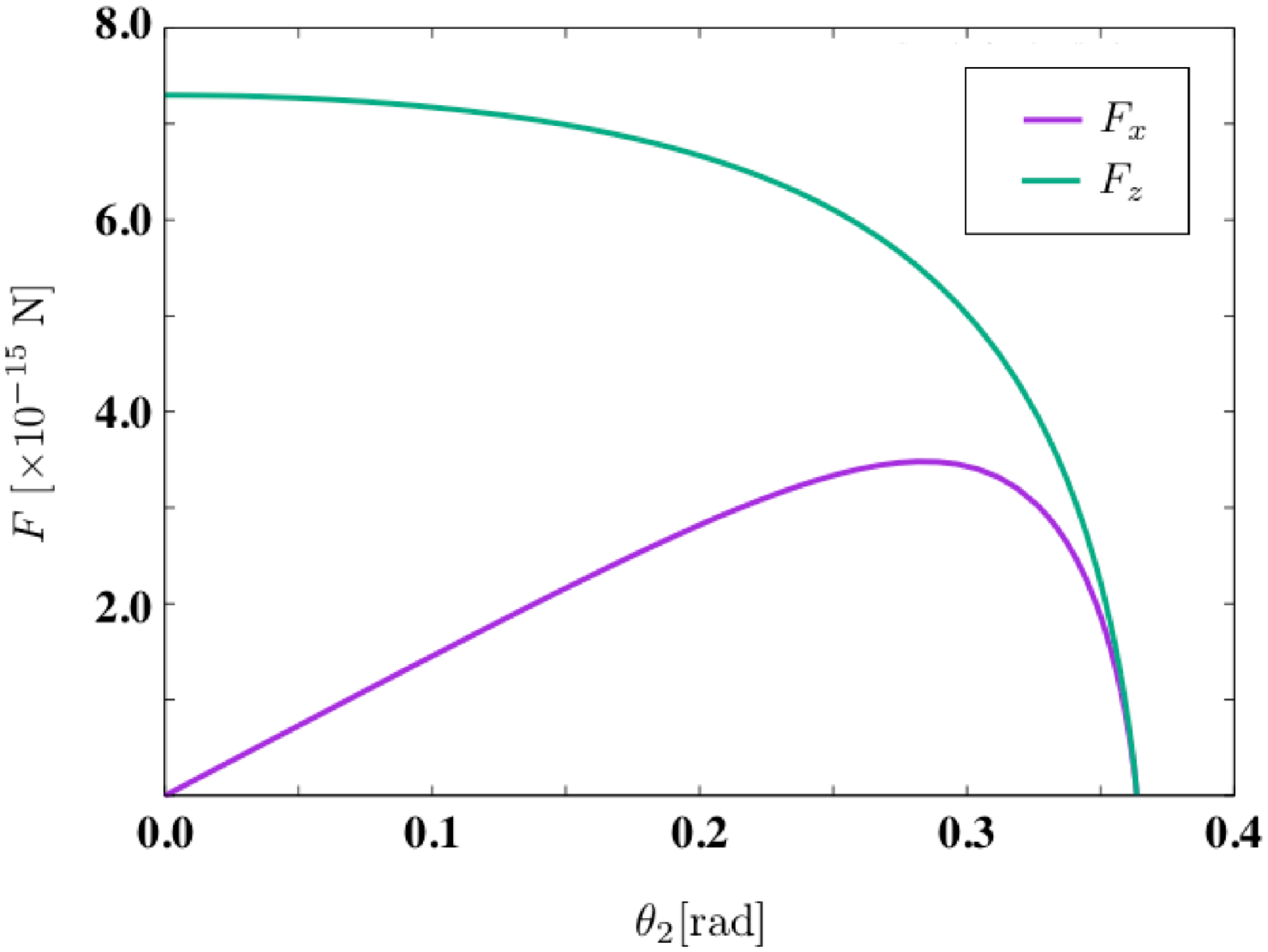}
  \end{minipage}
  \caption{
    Dissipation-assisted optical force. 
    The left figure is the plot of scattering radiation forces exerted on the gold nanosphere in the case without dissipation ($\tilde{n}_2 = 1.3+0i$). 
    The right top and bottom figures are the case with dissipation ($\tilde{n}_2 =1.3 + 0.1i$ and $ 1.3 + 1.3i$, respectively). 
    In these graphs, the purple lines indicate the radiation force in the $x$-direction $F_x$, and the green lines the radiation force in the $z$-direction $F_z$. 
    The excitation wavelength is 650 nm in all plots.
  }
  \label{fig:excess_opt_force}
\end{figure}

\par In FIG. \ref{fig:excess_opt_force}, we compare the scattering radiation force on a 100 nm gold nanosphere by 650 nm excitations with and without dissipation.
The left figure is the case without any dissipation ($\tilde{n}_2 = 1.3 + 0i$), and the right top and bottom figures are with dissipation ($\tilde{n}_2 = 1.3 + 0.1i$ and $1.3 + 1.3 i$, respectively).
It can be said that dissipation assists optical force in both $x$ and $z$ direction.
At large $\theta_2$, the radiation force reaches 0. This is because the larger angle of refraction is, the smaller Fresnel coefficient is.
When the angle of incidence is $\pi/2$, the transmission coefficient is zero.

\begin{figure}
  \begin{minipage}{0.5\hsize}
  \centering
  \includegraphics[width=8cm]{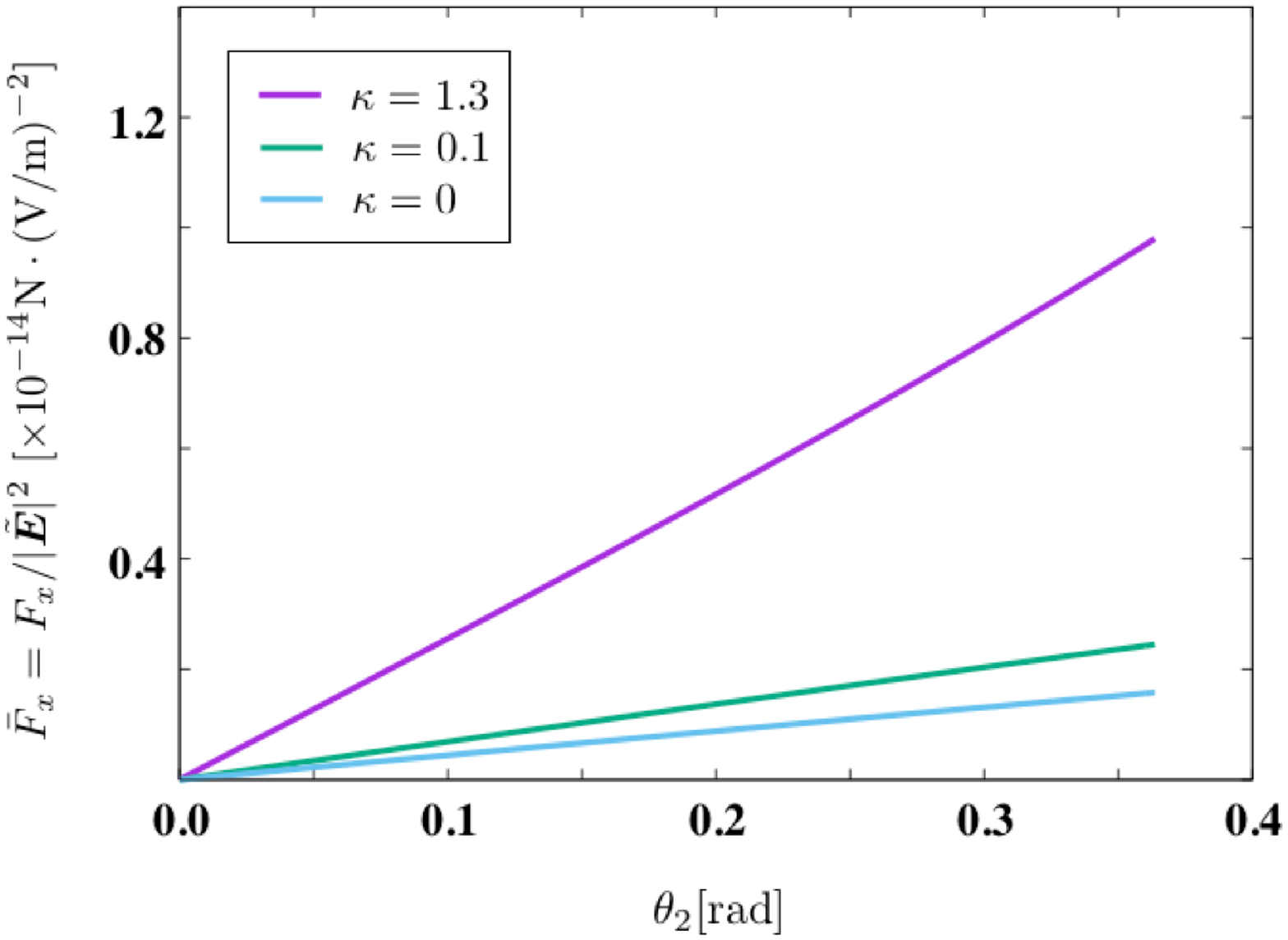}
  \end{minipage}
  \begin{minipage}{0.5\hsize}
  \centering
  \includegraphics[width=8cm]{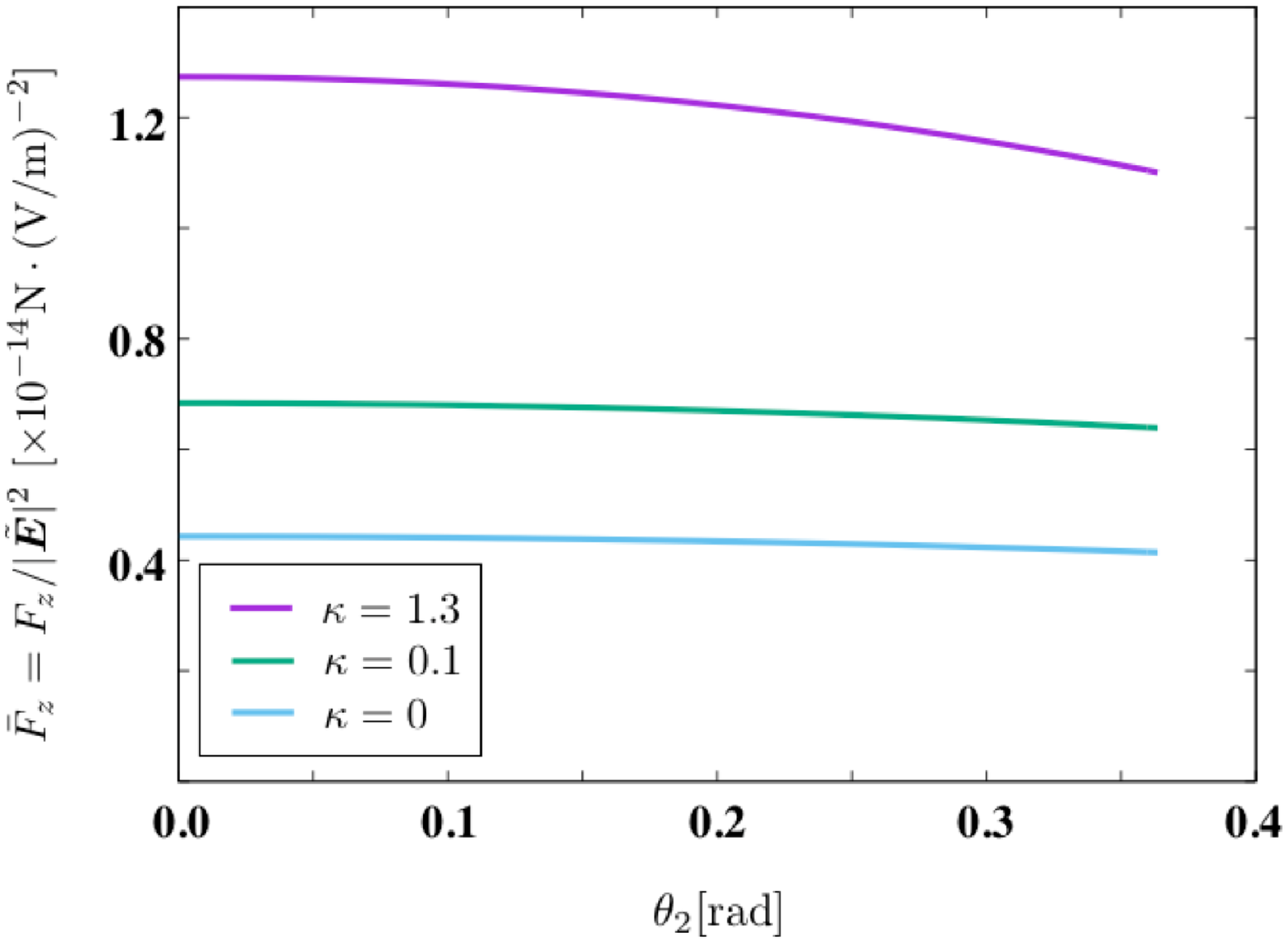}
  \end{minipage}
  \caption{
    Optical forces per unit intensity for various dissipation parameters ($\tilde{n}_2 = 1.3 + 0i, 1.3 + 0.1i, 1.3 + 1.3i$).
    The left figure shows the radiation force in the $x$-direction, and the right figure is the plot for the radiation force per in the $z$-direction.
    It is clear that in both $x$ and $z$ directions, the optical radiation force is assisted.
  }
  \label{fig:diss_nondiss}
\end{figure}

To clarify the excess wavenumber effect, we plot optical force on particle per unit power in FIG. \ref{fig:diss_nondiss} and compare the dissipative cases and the non-dissipation case.
We can confirm in both $x$ and $z$ directions optical force per intensity is assisted by dissipation.
The enhancement of the force which we can see at $\theta_2 = 0$ is a contribution from the polarisability of the probe particle.
The radiation force is not only proportional to the wavevector of the field but also to the polarisability of the probe particle, which increases with the imaginary part of the refractive index of surrounding media.
This is another factor of the the enhancement of the radiation force.

\begin{figure}[tbp]
  \centering
  \includegraphics[width=10cm]{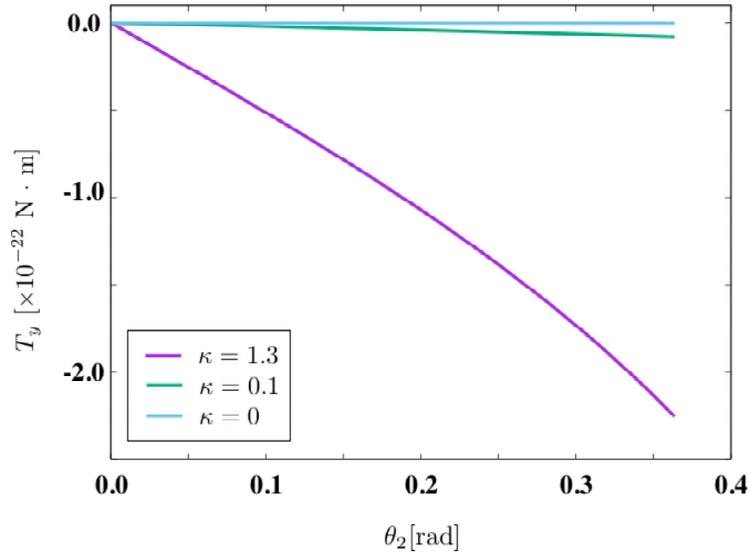}
  \caption{
    Transverse torques exerted by anomalous circular polarisation on 100 nm gold nanosphere. 
    The excitation wavelength is 650 nm. 
    The purple curve and the green curve are the $y$-components of the optical torques $T_y$ with dissipation ($\tilde{n}_2 = 1.3 + 1.3i$ and $1.3 + 0.1i$, respectively), while the blue curve is the torque without any dissipation ($\tilde{n}_2 = 1.3 + 0i$). 
    It is obvious that there is non-zero transverse torque with dissipation, whereas there is no torque without dissipation since the field does not rotate in the transverse direction without dissipation.
  }
  \label{fig:transverse_torque}
\end{figure}
\par FIG. \ref{fig:transverse_torque} shows transverse optical torques on the gold nanosphere induced by anomalous circular polarisation, which cannot be generated without dissipation.
As the anomalous circular polarisation vanishes at $\theta_2 \rightarrow 0$, the transverse optical torque vanishes.
This confirms that the anomalous polarisation causes the transverse torque.

\par The optical force and torque are proportional to the real part of the wavevector and the spin vector, respectively.
Both of these vectors become larger as the dissipation parameter increases.
Since there also exists Fresnel coefficient contribution, it cannot simply be said that the force and the torque increase with the dissipation parameter.
However, optical force and torque per unit intensity do monotonically increase with the dissipation.
The physical meaning of this statement is that large dissipation causes strong compression of light near the interface and results in enhancements of the momentum and the spin of photons.

\section{Conclusion}
To sum up, we utilised a complex-angle approach for calculating electromagnetic field near interfaces with dissipation, and revealed dissipation-induced extraordinary behaviours of the field: production of excess wavevector and generation of anomalous transverse circular polarisation. 
We also studied what kind of effects these behaviours cause on optical force and torque.
Excess wavevector assisted optical force, and anomalous circular polarisation generated transverse optical torque.
Since the effects discussed in this paper can be produced simply by adding dissipation, they are easy to explore in experiments. These effects add additional degrees of freedom for optical trapping and manipulation.

\ack
I thank Professor Hajime Ishihara and Professor Tomohiro Yokoyama for fruitful discussions on optical force and torque.
I thank Thomas Hodson, Samuel Palmer, and Yuki Kondo for proofreading and helping me to improve this manuscript.

\section*{Appendix: Polarisability of a small particle}
According to literatures \cite{bohren2008absorption, nieto2010optical, bekshaev2013subwavelength, bliokh2014extraordinary}, we can use the formula below to calculate the polarisability of a subwavelength spherial particle.
\begin{equation*}
  \alpha_e \simeq \frac{\varepsilon (\varepsilon_p - \varepsilon)}{\varepsilon_p + 2\varepsilon}a^3.
\end{equation*}
Here, $\varepsilon_p$ and $\varepsilon$ are the permittivity of the particle and that of the surrounding media, respectively. 
$a$ is the radius of the particle.
\section*{References}
\providecommand{\newblock}{}

\bibliographystyle{iopart-num}

\end{document}